\documentclass[letterpaper, 10 pt, conference]{ieeeconf}  

\IEEEoverridecommandlockouts                              

\usepackage{amsfonts, dsfont, mathtools, amsmath, amssymb, bbold, mathrsfs, bm, upgreek}
\usepackage{graphicx, float, epsfig, color, psfrag, adjustbox, enumerate, lipsum}
\usepackage[font=small, labelfont=bf]{caption}
\usepackage{refcount, url, cleveref}
\usepackage[table]{xcolor}
\usepackage{times, tabularx, subfigure, float} 
\usepackage{nicefrac, multirow, multicol, arydshln, multimedia}
\usepackage[noadjust]{cite}

\title{Passive elasticity properties of \textit{Octopus rubescens} arm}
\author{Udit Halder$^1$, Ekaterina Gribkova$^{1,2}$, Rhanor Gillette$^{2,3}$,  Prashant G. Mehta$^{1,4}$
\thanks{$^{1}$Coordinated Science Laboratory, $^{2}$Neuroscience Program, $^3$Department of Molecular and Integrative
Physiology, $^4$Department of Mechanical Science and Engineering, University of Illinois Urbana-Champaign, IL, 61801, USA
  Corresponding e-mail:  {\tt\small udit@illinois.edu}}%
\thanks{The authors gratefully acknowledge financial support from ONR MURI N00014-19-1-2373.}%
\thanks{The authors gratefully acknowledge Roddel Remy and Kathy Walsh for their help and advice for performing the elasticity experiments, and the Materials Research Laboratory at the University of Illiois Urbana-Champaign where the elasticity experiments were performed.}  
  }

\begin{document}
\bstctlcite{BSTcontrol} 
\maketitle
\thispagestyle{empty}
\pagestyle{empty}


\begin{abstract}
In this report, passive elasticity properties of \textit{Octopus rubescens} arm tissue are investigated using a multidisciplinary approach encompassing biomechanical experiments, computational modeling, and analyses. Tensile tests are conducted to obtain stress-strain relationships of the arm under axial stretch. Rheological tests are also performed to probe into dynamic shear response of the arm tissue. Based on these tests, comparisons against three different viscoelasticity models are reported.
\end{abstract}


\section{Introduction}
Flexible octopus arms have long fascinated researchers with their extraordinary ability to deform in all directions, including bending, extending, and twisting\,\cite{levy2017motor, hanlon2018cephalopod}. 
The goal of this report is to investigate arm elasticity which not only provides insights into the biomechanics of the arms but also holds implications for a broader spectrum of scientific disciplines, including  materials science and soft robotics~\cite{fung2013biomechanics, tramacere2014structure}. 

Octopus arm anatomy and movement patterns have been widely studied over the past few decades~\cite{nesher2020octopus, sumbre2001control}. Despite the many advances in experimental methods, systematic quantitative characterization of arm elasticity has been scarce until recently. In~\cite{tramacere2014structure}, mechanical properties of octopus arm suckers were reported, whereas elasticity properties of the arm musculature were studied in~\cite{di2021beyond, zullo2022octopus}. 
On the other hand, mathematical and computational modeling of slender flexible structures, and especially octopus arms, has seen an increased interest from roboticists~\cite{laschi2012soft, rus2015design, chang2020energy}. Many of these mathematical models~\cite{yekutieli2005dynamic, gazzola2018forward, chang2023energy} make use of nonlinear elasticity theory~\cite{antman1995nonlinear}, wherein specification of constitutive models of the elastic material becomes necessary. For example, the mechanical properties of specific muscle groups~\cite{di2021beyond, zullo2022octopus} have been beneficial in creating a detailed computational analog of an octopus arm~\cite{tekinalp2023topology}. 
 Furthermore, modeling the arm as a single flexible entity which is subject to internal actuation (muscles) has been shown to be productive in gaining deeper mathematical understanding of the arm~\cite{chang2023energy, chang2021controlling, wang2022sensory}.  
 All of these computational models create the need for a quantitative analysis of arm elasticity as a whole (as opposed to specific muscle groups or suckers). 

The primary contribution of this report is quantitative assessment of the passive elastic properties of octopus arm tissue through biomechanical experiments. Using \textit{Octopus rubescens} as a model species, dynamic mechanical analysis and rheological analysis are performed to obtain passive elasticity properties of the tissue. Tensile tests yield passive stress-strain curves, revealing the mechanical behavior of the material undergoing axial stretch. For better accuracy of the results, confocal laser scanning microscopy is used to measure arm sample geometries. Rheology tests provide insight into the shear modulus and viscoelastic properties of the material which are then compared against three viscoelasticity models.


\begin{figure*}[!t]
\centering
\includegraphics[width=\textwidth, trim = {0pt 140pt 0pt 120pt}, clip = true]{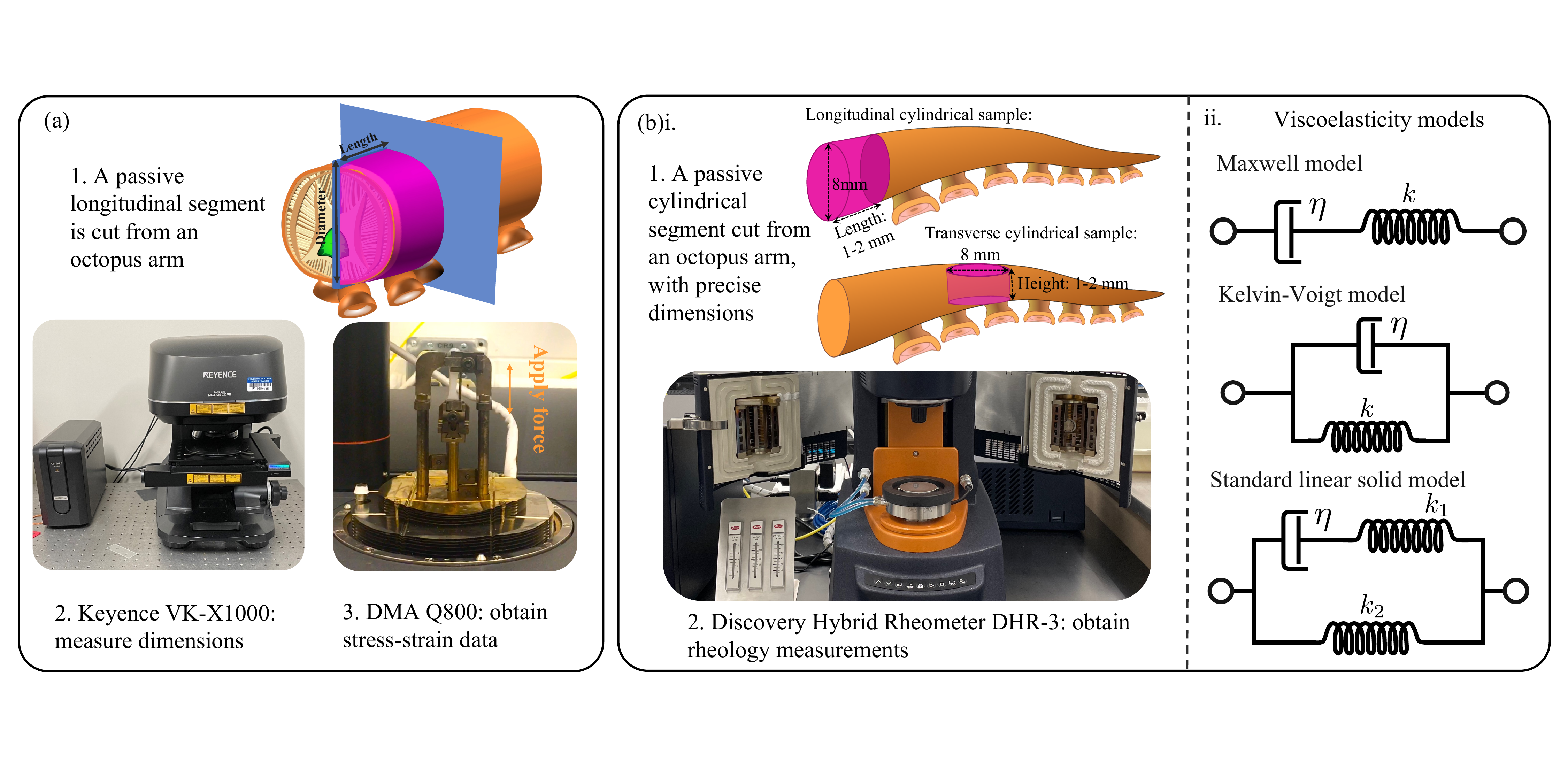}
\caption{Methods. (a) Process of obtaining arm samples, measuring their dimensions, and performing tensile tests. (b)i. Process of conducting rheological tests, and ii. three viscoelascity models. 
}
\label{fig:methods}
\vspace{-5pt}
\end{figure*}

\section{Methods} \label{sec:methods}

\subsection{Animal care} \label{sec:animal}

\noindent {\bf Animals.}
Specimens of \textit{Octopus rubescens} trapped in Monterey Bay, CA were purchased from Monterey Abalone Co. (Monterey, CA) and housed separately in artificial seawater (ASW) at 11-12$^\circ$C. Two animals (50g – 90g) were used in these experiments. Animals were fed pieces of shrimp or squid flesh every 1-3 days. All experiments were carried out in accordance with protocol \#23015 approved by the University of Illinois Urbana-Champaign (UIUC) Institutional Animal Care and Use Committee (IACUC).

\medskip
\noindent
{\bf Arm sectioning.}
An \textit{Octopus rubescens} was anesthetized in 2\% ethanol in chilled ASW, and an arm was isolated using a razor blade. For stress-strain measurements, longitudinal cylindrical or half-cylindrical samples were cut from the isolated arm using a razor blade, with minimum diameter of 2 mm (Fig.~\ref{fig:methods}a). For rheology measurements, cylindrical samples were cut from the isolated arm with specific dimensions: 8 mm diameter and 1-2 mm length (Fig.~\ref{fig:methods}b). Specifically, longitudinal cylindrical samples were cut simply using a razor blade, while transverse cylindrical segments were cut using a cork borer. All samples were placed into chilled 330 mM magnesium chloride solution to provide muscle relaxation.

\subsection{Arm segment measurements}
Accurate dimensional measurements of the severed octopus arm segments are required to obtain the elastic properties of the material. In particular, the cross-sectional area (denoted by $A_0$) of the cylindrical samples measured by traditional methods (e.g. using a digital caliper) are unreliable due to the softness of the material. Thus the segments were measured using a Keyence VK-X1000 3D Optical Profiler (Keyence Cooperation, Osaka, Japan) (Fig.~\ref{fig:methods}a). The segments were positioned on the center of the microscope stage and their surface morphologies were measured by confocal laser scanning microscopy. 


\subsection{Tensile tests} \label{sec:tensile_method}

After measuring the arm segments, tensile tests were performed to obtain stress-strain curves. The experiments were conducted using a DMA Q800 dynamic mechanical analyzer (Texas Instruments, Dallas, Texas, USA) (Fig.~\ref{fig:methods}a). For the stress-strain experiment, three main components of the instrument were of interest -- sample clamps, drive motor, and optical encoder. First, an arm segment whose geometry had already been measured, was mounted and secured between the top and bottom clamps of the instrument. The optical encoder was then used to measure the initial length of the sample ($L_0$). Next, the drive motor was used to apply equal and opposite forces ($F$) to the sample, causing it to elongate. The force was continually increased in small amounts yielding a stress on the sample $\sigma = \frac{F}{A_0}$.
The optical encoder was used to measure the corresponding length $L$ of the sample, resulting in the strain $\varepsilon = \frac{L-L_0}{L_0}$.

All tensile tests were performed in room temperature (23-25$^\circ$C). To avoid the decay of the arm tissue, samples were kept in chilled ASW, except during the measurement. 

\medskip
\noindent
{\bf Model.}
The stress-strain ($\sigma$-$\varepsilon$) curves are plotted to reveal mechanical properties of the sample. 
Polynomials of order $n$ are used to characterize the stress-strain relationship
\begin{align}
    \sigma = \sum_{i=1}^n a_i \varepsilon^i,
    \label{eq:pol_fit}
\end{align}
where the coefficients $a_i$ are obtained using linear regression. In particular, the slope of this curve indicates the modulus of elasticity or Young's modulus in the linear (Hookean) region~\cite{meyers2008mechanical}. 

\subsection{Rheology tests} \label{sec:rheology_method}
Rheological experiments were performed using a Discovery Hybrid Rheometer (DHR-3, Texas Instruments, Dallas, Texas, USA), as shown in Fig.~\ref{fig:methods}(b)i. As indicated in \S\,\ref{sec:methods}\ref{sec:animal}, both longitudinally and transversely cut cylindrical arm sections (of diameter 8~mm) were used for the rheology experiment. The samples were placed between two parallel plates of diameter 8~mm. Next, torques of various frequencies were applied between the two plates by the rheometer so that the sample's top and bottom surfaces undergo a shear. The instrument then measured both the strain and stress on the sample. Temperature was kept constant at 25$^\circ$C during all rheological experiments. 

\begin{figure*}[!t]
\centering
\includegraphics[width=1.0\textwidth, trim = {15pt 45pt 25pt 10pt}, clip = true]{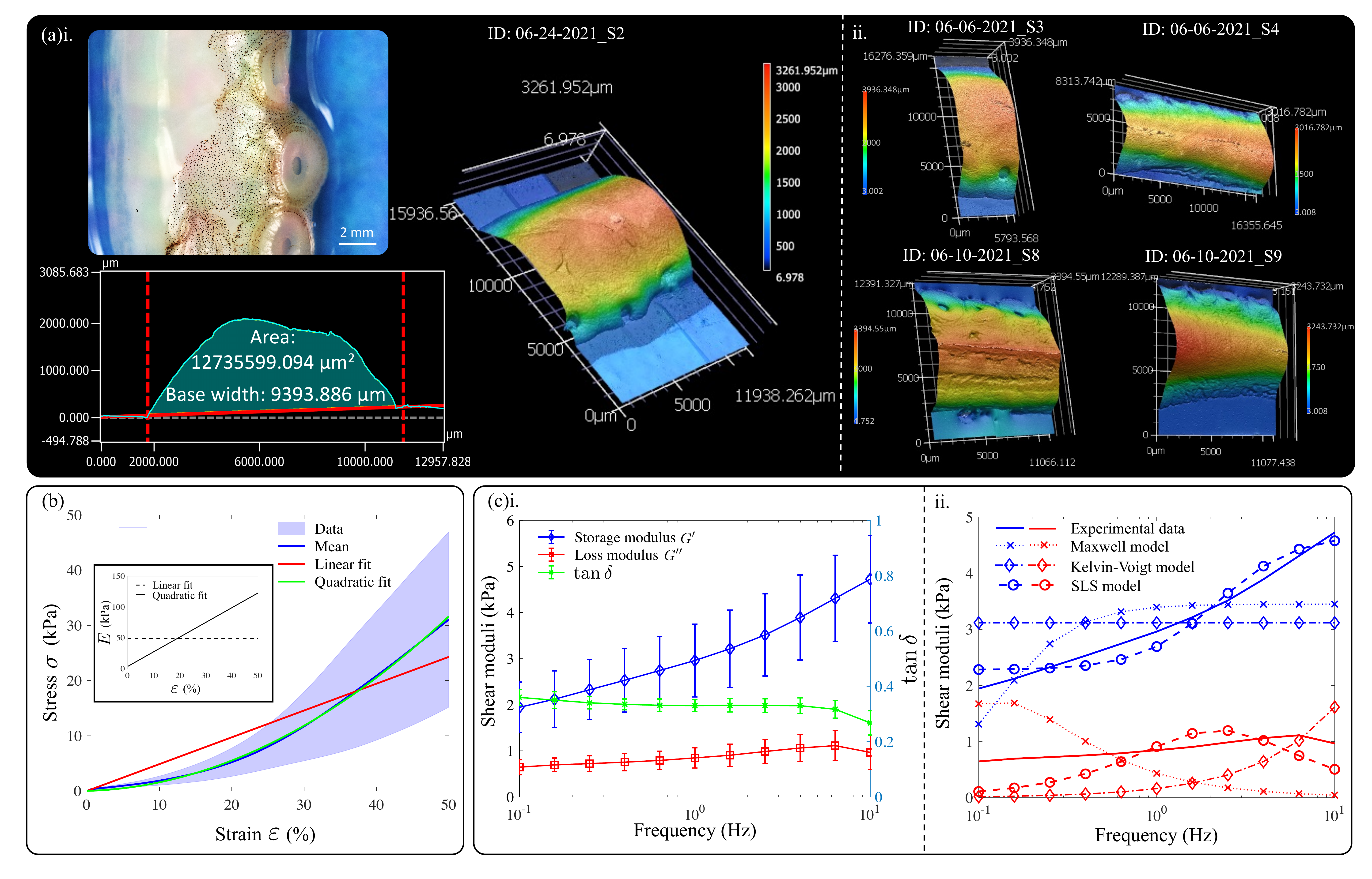}
\caption{Results. (a)i. Arm cross-sectional area measurements: the sample is shown in the top left, its 3D surface reconstruction is shown in the right, and a cross-sectional morphology is shown in the bottom left. (ii) 3D surface reconstructions are shown for four other samples. (b) Tensile test: stress-strain data $\pm$ std are shown in blue (mean in dark blue) with polynomial fits -- linear (red) and quadratic (green). Slopes of the polynomials (Young's modulus $E$) are shown in the inset. (c) Rheology experiment: i. dynamic moduli (storage modulus $G'$ in blue, loss modulus $G''$ in red) and $\tan \delta$ (in green) are plotted as a function of oscillation frequency, and ii. fits of the three viscoelasticity models ($G'$ in blue and $G''$ in red). 
}
\label{fig:results}
\vspace{-5pt}
\end{figure*}

\medskip
\noindent
{\bf Model.}
Let the strain and stress of the sample be represented as $\varepsilon (t) = \varepsilon_0 \sin (\omega t)$ and $\sigma (t) = \sigma_0 \sin (\omega t + \delta)$, respectively, where $\varepsilon_0$ and $\sigma_0$ are the peak strain and stress, respectively; $\omega$ is the frequency of the applied torque oscillation; and $\delta = \delta (\omega)$ is the phase lag between stress and strain. Then the shear storage ($G'$) and loss ($G''$) moduli are calculated as~\cite{meyers2008mechanical}
\begin{align}
     G'(\omega) = \frac{\sigma_0}{\varepsilon_0} \cos (\delta (\omega)), \quad G'' (\omega) = \frac{\sigma_0}{\varepsilon_0} \sin (\delta (\omega)).
\end{align}
The shear storage modulus ($G'$) is indicative of the shear modulus of the arm tissue, whereas the shear loss modulus ($G''$) provides insights into the viscoelastic properties of the material.   

To assess the viscoelastic properties, three different viscoelasticity models are considered -- the Maxwell (M) model, the Kelvin-Voigt (K-V) model, and the standard linear solid (SLS) model~\cite{fung2013biomechanics, christensen2012theory}. The basic mechanical units used in these models are the elastic spring (elasticity modulus $k$) and viscous dashpot (viscosity coefficient $\eta$), with governing consitutive equations $\sigma = k \varepsilon$ and $\sigma = \eta \dot{\varepsilon}$, respectively ($\dot{\varepsilon}$ denotes the strain rate). These elements are then arranged in series (Maxwell) or parallel (Kelvin-Voigt) or a combination of both (SLS), as illustrated in Fig.~\ref{fig:methods}(b)ii. The dynamic moduli of these three models are given as follows~\cite{bonfanti2020fractional}
\begin{align}
\begin{split}
    &G'_{\text{M}} (\omega) = \frac{\eta^2 k \omega^2}{k^2 + \eta^2 \omega^2}, ~~~  
    G''_{\text{M}} (\omega) = \frac{\eta k^2 \omega}{k^2 + \eta^2 \omega^2}, \\
    &G'_{\text{K-V}} (\omega) = k, ~~~ G''_{\text{K-V}} (\omega) = \eta \omega, \\
    &G'_{\text{SLS}} (\omega) = \frac{k_1^2 k_2 + \eta^2 \omega^2 (k_1+k_2)}{k_1^2 + \eta^2 \omega^2}, ~~~  
    G''_{\text{SLS}} (\omega) = \frac{\eta k_1^2 \omega}{k_1^2 + \eta^2 \omega^2}. 
\end{split}
\label{eq:moduli_model}
\end{align}
The parameters for each model include the coefficients of elasticity and viscosity, which were obtained by solving a least squares problem to fit the experimentally obtained data. 

\section{Results} \label{sec:results}

\subsection{Arm segment measurements}
For tensile stress-strain measurements, arm sample scan data from Keyence VK-X1000 3D Optical Profiler was analyzed using Keyence MultiFileAnalyzer software v2.1.2.17. A total of nine samples (see details in Table~\ref{tab:keyence}) were used and at least four evenly-spaced cross-section measurements were taken for each sample to obtain an average cross-sectional area. For each measurement, a transverse profile line was drawn across the sample surface scan, producing a surface trace (see Fig.~\ref{fig:results}(a)i.). Then an upper cross-sectional area measurement was taken between two points on the surface trace that indicated the bottom left- and right-most edges of the sample. In addition, the distance between these two points was also measured to calculate the average base widths for each sample. 

\begin{table}[t]
	\small
	\centering
	\captionof{table}{Arm segment area measurements}
	\begin{tabular}{cccc}
		\rowcolor{violet}
		{} & {} & {\color{white} Average area} & {\color{white} Average base} \\
            \rowcolor{violet}
             \multirow{-2}{*}{\color{white} Sample ID} & \multirow{-2}{*}{\color{white} Location} &{\color{white} $A_0$ (mm$^2$)} & {\color{white} width (mm)} \\
		\hline\noalign{\smallskip}
		06-06-2021\_S3 & medial & 17.032 & 13.057 \\ 
            06-06-2021\_S4 & distal & 06.618 & 7.347 \\ 
            06-10-2021\_S3 & proximal & 18.436 & 11.912 \\ 
            06-10-2021\_S4 & medial & 26.171 & 10.005 \\ 
            06-10-2021\_S6 & medial & 12.230 & 9.429 \\ 
            06-10-2021\_S8 & distal & 14.756 & 9.280 \\ 
            06-10-2021\_S9 & distal & 21.301 & 8.585 \\ 
            06-24-2021\_S2 & proximal & 16.142 & 9.545 \\ 
            06-24-2021\_S3 & medial & 12.656 & 8.937 \\ 
		\hline
	\end{tabular}
	\label{tab:keyence}
	       \vspace*{-10pt}
\end{table}

\subsection{Tensile test results}
The tensile stress-strain data of all the samples are plotted in Fig.~\ref{fig:results}b. The data showed high variability especially at high strains. For this reason, data up to 50\% strain are plotted with the mean plotted in blue. In addition to the raw data, polynomial fits (equation~\eqref{eq:pol_fit}) of the orders $n=1$ (in red) and $n=2$ (in green) are also plotted. The coefficients of the fitted polynomials are (also given in Table~\ref{tab:parameters}) $a_1 = 48.559$ kPa for the linear fit ($n=1$) and $a_1 = 3.678$ kPa, $a_2 = 119.084$ kPa for the quadratic fit ($n=2$). As is seen from Fig.~\ref{fig:results}b, the quadratic fit matches the mean stress-strain curve. The slopes of the polynomial curves are then obtained as $a_1$ for linear fit and $a_1 + 2a_2 \varepsilon$ for quadratic fit, from which the effective Young's modulus ($E$) can be calculated for a given strain and is reported in the inset of Fig.~\ref{fig:results}b. The effective Young's modulus remains in the 1-150 kPa range for strains $< 50 \%$, consistent with other reported results on soft biological materials~\cite{samani2007elastic, van2013contact, tramacere2014structure}.    

\begin{table}[t]
	\normalsize
	\centering
	\captionof{table}{Parameters for model-fitting.  Stress-strain model parameters are defined in~\eqref{eq:pol_fit} and viscoelasticity model parameters are defined in~\eqref{eq:moduli_model} (here $k=k_1$ for SLS model). }
	\begin{tabular}{c|ccc}
		\rowcolor{violet}
            \multicolumn{4}{c}{\color{white} Stress-strain models}  \\
            \rowcolor{violet!60}
		\multicolumn{2}{c|}{\color{white} Parameters} & {\color{white} Linear} & {\color{white} Quadratic} \\
            \multicolumn{2}{c|}{$a_1$ [kPa]} & 48.559 & 3.678\\
            \multicolumn{2}{c|}{$a_2$ [kPa]} & -- &  119.084 \\
            \hline
            \rowcolor{violet}
            \multicolumn{4}{c}{\color{white} Viscoelasticity models}  \\
            \rowcolor{violet!60}
            {\color{white} Parameters} & {\color{white} Maxwell} & {\color{white} Kelvin-Voigt} & {\color{white} SLS}\\
            $\eta$ [kPa$\cdot$s] & 4.299 & 0.026 & 0.175\\
            $k$   [kPa]          & 3.449 & 3.116 & 2.412\\
            $k_2$ [kPa]          & --   & --     & 2.278\\
             \hline
	\end{tabular}
	\label{tab:parameters}
	        \vspace*{-10pt}
\end{table}

\vspace*{-0pt}
\subsection{Rheology test results}
The results of the rheology experiments are graphically illustrated in Fig.~\ref{fig:results}c. Experimental data for frequencies up to 10~Hz are reported because of the unreliability of the data for higher frequencies.
As is seen from Fig.~\ref{fig:results}(c)i, both the storage~($G'$) and loss ($G''$) moduli show upward trends as the frequency of oscillation is increased, however the increment in $G''$ is less than that of $G'$. On the other hand, the tangent of the loss angle ($\tan \delta$) shows a slightly downward trend with increasing frequency.

Finally, comparisons against the three viscoelastic models (as described in \S\,\ref{sec:methods}\ref{sec:rheology_method}) are provided in Fig.~\ref{fig:results}(c)ii. The experimental means of both the storage and loss moduli are used to fit the models. The MATLAB function \textit{fminunc} is used to solve the least squares problem to obtain optimal parameter values for each model which are given as follows (also see Table~\ref{tab:parameters}): $\eta = 4.299$ kPa$\cdot$s, $k = 3.449$ kPa for the Maxwell model; $\eta = 0.026$ kPa$\cdot$s, $k = 3.116$ kPa for the Kelvin-Voigt model; and $\eta = 0.175$ kPa$\cdot$s, $k_1 = 2.412$ kPa, $k_2 = 2.278$ kPa for the SLS model. As is seen from Fig.~\ref{fig:results}(c)ii, the SLS model yields closest match with the experimental data. 

\section{Discussion} \label{sec:conclusion}
This report presents experimental methods, data, and model fitting to obtain passive elasticity properties of \textit{O. rubescens} arm tissue. 
Passive stress-strain data are obtained from tensile tests and arm area measurements. Polynomial model fitting reveals a quadratic relationship between passive stress and strain. Rheological experiments yield measurements of dynamic shear moduli and viscoelastic response of octopus arm tissue. Model fitting against three canonical viscoelasticity models demonstrates close match with a three-element standard linear solid model. 


The results of this study provide a holistic assessment of the passive elasticity properties of octopus arms that is important for several biophysical models~\cite{yekutieli2005dynamic, chang2023energy}. As a future work, an informed modeling of arm  passive elasticity will be considered, which will then be coupled with detailed modeling of the active components (muscles)~\cite{di2021beyond, zullo2022octopus}. Modeling of this kind will not only help advance our knowledge of cephalopod biology but will also inspire innovations in soft robotics and materials science~\cite{rus2015design}.


%
%
%


\bibliographystyle{IEEEtran}
\bibliography{references}

\begin{thebibliography}{10}
\providecommand{\url}[1]{#1}
\csname url@samestyle\endcsname
\providecommand{\newblock}{\relax}
\providecommand{\bibinfo}[2]{#2}
\providecommand{\BIBentrySTDinterwordspacing}{\spaceskip=0pt\relax}
\providecommand{\BIBentryALTinterwordstretchfactor}{4}
\providecommand{\BIBentryALTinterwordspacing}{\spaceskip=\fontdimen2\font plus
\BIBentryALTinterwordstretchfactor\fontdimen3\font minus
  \fontdimen4\font\relax}
\providecommand{\BIBforeignlanguage}[2]{{%
\expandafter\ifx\csname l@#1\endcsname\relax
\typeout{** WARNING: IEEEtran.bst: No hyphenation pattern has been}%
\typeout{** loaded for the language `#1'. Using the pattern for}%
\typeout{** the default language instead.}%
\else
\language=\csname l@#1\endcsname
\fi
#2}}
\providecommand{\BIBdecl}{\relax}
\BIBdecl

\bibitem{levy2017motor}
G.~Levy, N.~Nesher, L.~Zullo, and B.~Hochner, ``Motor control in soft-bodied
  animals: the octopus,'' in \emph{The Oxford Handbook of Invertebrate
  Neurobiology}, 2017.

\bibitem{hanlon2018cephalopod}
R.~T. Hanlon and J.~B. Messenger, \emph{Cephalopod behaviour}.\hskip 1em plus
  0.5em minus 0.4em\relax Cambridge University Press, 2018.

\bibitem{fung2013biomechanics}
Y.-c. Fung, \emph{Biomechanics: mechanical properties of living tissues}.\hskip
  1em plus 0.5em minus 0.4em\relax Springer Science \& Business Media, 2013.

\bibitem{tramacere2014structure}
F.~Tramacere, A.~Kovalev, T.~Kleinteich \emph{et~al.}, ``Structure and
  mechanical properties of octopus vulgaris suckers,'' \emph{Journal of The
  Royal Society Interface}, vol.~11, no.~91, p. 20130816, 2014.

\bibitem{nesher2020octopus}
N.~Nesher, G.~Levy, L.~Zullo, and B.~Hochner, ``Octopus motor control,''
  \emph{Oxford Research Encyclopedia of Neuroscience. Oxford University Press,
  USA}, 2020.

\bibitem{sumbre2001control}
G.~Sumbre, Y.~Gutfreund, G.~Fiorito \emph{et~al.}, ``Control of octopus arm
  extension by a peripheral motor program,'' \emph{Science}, vol. 293, no.
  5536, pp. 1845--1848, 2001.

\bibitem{di2021beyond}
A.~Di~Clemente, F.~Maiole, I.~Bornia, and L.~Zullo, ``Beyond muscles: role of
  intramuscular connective tissue elasticity and passive stiffness in octopus
  arm muscle function,'' \emph{Journal of Experimental Biology}, vol. 224,
  no.~22, p. jeb242644, 2021.

\bibitem{zullo2022octopus}
L.~Zullo, A.~Di~Clemente, and F.~Maiole, ``How octopus arm muscle contractile
  properties and anatomical organization contribute to arm functional
  specialization,'' \emph{Journal of Experimental Biology}, vol. 225, no.~6, p.
  jeb243163, 2022.

\bibitem{laschi2012soft}
C.~Laschi, M.~Cianchetti, B.~Mazzolai \emph{et~al.}, ``Soft robot arm inspired
  by the octopus,'' \emph{Advanced robotics}, vol.~26, no.~7, pp. 709--727,
  2012.

\bibitem{rus2015design}
D.~Rus and M.~T. Tolley, ``Design, fabrication and control of soft robots,''
  \emph{Nature}, vol. 521, no. 7553, pp. 467--475, 2015.

\bibitem{chang2020energy}
H.-S. Chang, U.~Halder, C.-H. Shih \emph{et~al.}, ``Energy shaping control of a
  cyberoctopus soft arm,'' in \emph{2020 59th IEEE Conference on Decision and
  Control (CDC)}.\hskip 1em plus 0.5em minus 0.4em\relax IEEE, 2020, pp.
  3913--3920.

\bibitem{yekutieli2005dynamic}
Y.~Yekutieli, R.~Sagiv-Zohar, R.~Aharonov \emph{et~al.}, ``Dynamic model of the
  octopus arm. i. biomechanics of the octopus reaching movement,''
  \emph{Journal of neurophysiology}, vol.~94, no.~2, pp. 1443--1458, 2005.

\bibitem{gazzola2018forward}
M.~Gazzola, L.~Dudte, A.~McCormick, and L.~Mahadevan, ``Forward and inverse
  problems in the mechanics of soft filaments,'' \emph{Royal Society Open
  Science}, vol.~5, no.~6, p. 171628, 2018.

\bibitem{chang2023energy}
H.-S. Chang, U.~Halder, C.-H. Shih \emph{et~al.}, ``Energy-shaping control of a
  muscular octopus arm moving in three dimensions,'' \emph{Proceedings of the
  Royal Society A}, vol. 479, no. 2270, p. 20220593, 2023.

\bibitem{antman1995nonlinear}
S.~S. Antman, \emph{Nonlinear Problems of Elasticity}.\hskip 1em plus 0.5em
  minus 0.4em\relax Springer, 1995.

\bibitem{tekinalp2023topology}
A.~Tekinalp, N.~Naughton, S.-H. Kim \emph{et~al.}, ``Topology, dynamics, and
  control of an octopus-analog muscular hydrostat,'' \emph{arXiv preprint
  arXiv:2304.08413}, 2023.

\bibitem{chang2021controlling}
H.-S. Chang, U.~Halder, E.~Gribkova \emph{et~al.}, ``Controlling a cyberoctopus
  soft arm with muscle-like actuation,'' in \emph{2021 60th IEEE Conference on
  Decision and Control (CDC)}.\hskip 1em plus 0.5em minus 0.4em\relax IEEE,
  2021, pp. 1383--1390.

\bibitem{wang2022sensory}
T.~Wang, U.~Halder, E.~Gribkova \emph{et~al.}, ``A sensory feedback control law
  for octopus arm movements,'' in \emph{2022 IEEE 61st Conference on Decision
  and Control (CDC)}.\hskip 1em plus 0.5em minus 0.4em\relax IEEE, 2022, pp.
  1059--1066.

\bibitem{meyers2008mechanical}
M.~A. Meyers and K.~K. Chawla, \emph{Mechanical behavior of materials}.\hskip
  1em plus 0.5em minus 0.4em\relax Cambridge university press, 2008.

\bibitem{christensen2012theory}
R.~Christensen, \emph{Theory of viscoelasticity: an introduction}.\hskip 1em
  plus 0.5em minus 0.4em\relax Elsevier, 2012.

\bibitem{bonfanti2020fractional}
A.~Bonfanti, J.~L. Kaplan, G.~Charras, and A.~Kabla, ``Fractional viscoelastic
  models for power-law materials,'' \emph{Soft Matter}, vol.~16, no.~26, pp.
  6002--6020, 2020.

\bibitem{samani2007elastic}
A.~Samani, J.~Zubovits, and D.~Plewes, ``Elastic moduli of normal and
  pathological human breast tissues: an inversion-technique-based investigation
  of 169 samples,'' \emph{Physics in medicine \& biology}, vol.~52, no.~6, p.
  1565, 2007.

\bibitem{van2013contact}
J.~Van~Kuilenburg, M.~A. Masen, and E.~Van Der~Heide, ``Contact modelling of
  human skin: What value to use for the modulus of elasticity?''
  \emph{Proceedings of the institution of mechanical engineers, Part J: Journal
  of Engineering Tribology}, vol. 227, no.~4, pp. 349--361, 2013.

\end{thebibliography}

\end{document}